**Feature**

# Enabling human-centered AI: A new junction and shared journey between AI and HCI communities

*Collaboration between the two communities will effectively enable human-centered AI*

**Wei Xu, Marvin Dainoff**

Artificial intelligence (AI) has brought benefits, but it may also cause harm if it is not appropriately developed. Current development is mainly driven by a "technology-centered" approach, causing many failures [1, 2]. For example, the AI Incident Database has documented over a thousand AI-related accidents [3]. To address these challenges, a human-centered AI (HCAI) approach has been promoted and has received a growing level of acceptance over the last a few years [1, 2]. HCAI calls for combining AI with HCI design will enable the development of AI systems (e.g., autonomous vehicles, intelligent user interfaces, or intelligent decision-making systems) to achieve its design goals such as usable/explainable AI, human-controlled AI, and ethical AI.

While HCAI promotion continues, it has not specifically addressed the collaboration between AI and HCI communities, resulting in uncertainty about what action should be taken by both sides to apply HCAI in developing AI systems [2].

This article focuses on the collaboration between the AI and HCI communities, which leads to nine recommendations for effective collaboration to enable HCAI in developing AI systems.

**History is repeating now**

When personal computers (PCs) emerged in the 1980s, their development followed a technology-centered approach, which ignored the needs of ordinary users and resulted in UX problems. With an initiative primarily driven by a collaboration by the professionals of human factors, computer science, and psychology, a new field of HCI emerged. Such an interdisciplinary collaborative field has benefited the promotion of a human-centered approach through user experience (UX)-driven design.

History now seems to be repeating, having brought the AI and HCI communities to a new juncture. This time the consequences of ignoring the human-centered approach are much more serious [3].

**New opportunities to begin a journey of collaboration**

AI technology has unique characteristics compared to non-AI systems [4]. The AI communities have recently made huge investments in the ethics of AI systems, acknowledging the value of identifying human-centered principles that can guide how AI systems are deployed. The HCI community has also been applying the human-centered approach to computing systems but faces some challenges in applying it to AI systems [5, 6]. How can the AI and HCI communities work together to develop human-centered AI systems?

To better understand the current transition of from traditional human interaction with non-AI systems to interaction with AI systems, we conducted a study to assess the transition between the two types of interaction under seven aspects (see the selected results in Table): (1) identify the major transitions from human interaction with non-AI to AI systems; (2) identify the unique characteristics introduced by AI technology; (3) analyze the pros and cons of the AI and HCI approaches to enable HCAI [4].

**The complementarity of approaches across the AI and HCI communities for enabling HCAI**

| Transitions | New Characteristics Introduced by AI Technologies | Pros and Cons of AI Approaches for Enabling HCAI | Pros and Cons of HCI Approaches for Enabling HCAI |
|---|---|---|---|
| From expected to unexpected machine behavior | • Unique autonomous capabilities (e.g., self-learning, adaptation, self-execution) may handle some scenarios that designers cannot anticipate<br>• AI systems exhibit unique behaviors and indeterministic outputs, causing potentially biased outcomes | Pros<br>• Handle some unexpected scenarios in emergencies<br>• Enable systems to be more productive based on new technologies (e.g., big data, machine learning) than individual humans<br>Cons<br>• Lack methods informed by behavioral science for managing machine behavior<br>• Do not consider evolving system behavior for testing/validation of software in current software engineering approach | Pros<br>• Conduct iterative design prototyping and testing activities to help data collection, training, and testing for optimizing algorithms<br>• Collect system and user feedback to support continued improvement of machine behavior<br>• Deploy user-participatory design to support human-centered & interactive machine learning<br>Cons<br>• Design relies on anticipated system behaviors and cannot deal with abnormal/unexpected scenarios |
| From interaction to potential human-AI collaboration | • Intelligent agents with the autonomous capabilities may partner with humans as teammates for collaboration | Pros<br>• Develop intelligent agents to work as teammates with humans<br>• Enhance overall system performance through human-AI collaboration<br>Cons<br>• Lack cognitive models and theories to implement human-AI collaboration<br>• Primarily adopt a technology-centered approach (i.e., humans adapt to AI) | Pros<br>• Leverage existing human-human theories for modeling human-machine collaboration<br>• Help measure human-AI collaboration with current HCI evaluation methods<br>• Primarily adopt a human-centered approach<br>Cons<br>• Lack mature models for human-AI collaboration (e.g., mutual trust and situation awareness, shared control)<br>• Focus on human interaction with non-AI systems in current approaches<br>• Consider machines as assistive tools (not collaborative teammates) in current approaches |
| From siloed human intelligence | • Machines possess some cognitive capabilities like humans (e.g., learning, | Pros<br>• Develop hybrid intelligence to produce more powerful intelligence by taking the | Pros<br>• Advocate human-controlled hybrid intelligence<br>• Potential to help modeling advanced human |
| to hybrid intelligence | reasoning) | complementary advantages of human and machine intelligence<br>Cons<br>• Hard to emulate advanced human cognitive capabilities, resulting in challenges in developing machine intelligence in isolation<br>• May not guarantee human-centered design in hybrid intelligence (i.e., humans may loss ultimate decision-making authority) | cognitive capabilities<br>Cons<br>• Lack methods for modeling advanced human cognition<br>• Lack consideration of machine intelligence in system design |
| From user interface usability to AI explainability (XAI) | • "AI black box" effect may make the output obscure to users, impacting user trust and acceptance | Pros<br>• Develop transparent and explainable algorithms<br>• Develop explainable UI models with advanced HCI technology<br>Cons<br>• Generate more complex algorithms with an algorithm-driven approach, resulting in less explainable AI<br>• Lack participation from other disciplines<br>• Lack validated psychological models for explanation | Pros<br>• Help deliver XAI with advanced UI and visualization technologies<br>• Help accelerate the transfer of psychological theories<br>• Build human-centered XAI through user-exploratory and user-participation approaches<br>• Help validate XAI with HCI testing methods<br>Cons<br>• Focus on system usability in current HCI approaches |
| From human-centered automation to human-controlled autonomy | • Different from automation (based on fixed rules or logics), autonomous systems may produce indeterministic results and can handle some unexpected situations<br>• Public's perception and over-trust of autonomous technology may cause safety issues<br>• Confusions between automation and autonomy | Pros<br>• Develop autonomous systems to handle some unexpected scenarios in emergencies that could not be done using automation<br>Cons<br>• Develop systems that may be without ultimate human control due to lack of HCAI approach | Cons<br>• Transfer the lessons learned from human factors/HCI research on automation to avoid issues (e.g., over-trust, human "out-of-the-loop")<br>• Develop innovative research/design paradigm (e.g., collaborative human-autonomy teaming)<br>Cons<br>• Lack effective design paradigm that can enable human operators to monitor and quickly take over control of autonomous systems in emergencies |
| | may lead to inappropriate expectations and potential misuse of technology | | |
| From conventional interactions to intelligent interactions | • Different from traditional the "stimulus-response" paradigm, intelligent systems can initiate proactive interaction<br>• Paradigmatic changes emerge for interaction design, e.g., user intent detection, affective interaction, "fuzzy reasoning" interaction | Pros<br>• Develop intelligent interaction technologies with new technologies (e.g., AI, big data, sensing technologies)<br>Cons<br>• Lack effective design paradigms and metaphors for intelligent interactions<br>• Cause overload of human cognitive resource in pervasive computing environment (e.g., ambient intelligence) | Pros<br>• Drive effective design paradigms and metaphors for intelligent interactions<br>• Design/develop natural and usable intelligent UI<br>• Adapt AI technology to human capabilities by optimizing the match of human cognitive workload and interaction technology<br>Cons<br>• Focus on traditional interaction design in current HCI design standards<br>• Late participation of HCI professionals in the development of intelligent systems<br>• Cannot effectively support the development of intelligent interactions with current HCI design & prototyping methods |
| From general user needs to specific ethical AI needs | • New user needs are emerging, e.g., privacy, fairness, ethics | Pros<br>• Increased awareness of ethical AI issues in developing AI systems<br>• Many ethical AI guidelines published<br>Cons<br>• Lack effective methods to translate ethical AI principles into practice<br>• Eethical design may be ad-hoc activities after development<br>• Lack technical examples for ethical design<br>• AI engineers typically lack formal training in applying ethics to design and tend to view ethical deign as another form of solving technical problems | Pros<br>• Leverage methods (e.g., iterative design and testing) to translate user needs into the process of data collection as well as training, optimizing, and testing of algorithms and machine behaviors<br>• Leverage HCI interdisciplinary skills to address ethical AI issues by adopting social and behavioral science methods<br>Cons<br>• Lack best known methods for addressing ethical AI |

As the table above shows, there are unique characteristics introduced by AI technology, posing challenges in developing HCAI systems. The approaches taken by each community show their individual work focus but more importantly, both sides reveal the gaps in developing HCAI systems.

However, the complementarity nature of the AI and HCI communities in term of approaches encourages a collaboration between both communities, which will lead to developing human-centered AI systems more effectively. For instance, the HCAI design goal is to develop human-controlled AI. Driven by the HCAI approach, we advocate that hybrid intelligence must be developed in a context of "human machine" systems by leveraging the complementary advantages of AI and human intelligence to produce a more powerful intelligence form: human-machine hybrid intelligence. This strategy not only solves the bottleneck effect of developing AI technology in isolation, but also emphasizes the use of humans and machines as a system (human-machine system) and integrates human functions and roles into AI systems as the ultimate decision makers without harming humans. As an interdisciplinary, future HCI work needs to help the AI community explore cognitive computing based on human cognitive abilities (e.g., intuitive reasoning, knowledge evolution) in support of developing human-machine hybrid intelligent systems. Future HCI work should help accelerate the conversion of existing psychological research results to support the work of cognitive computing and define cognitive architecture for AI research. HCI professionals also need to collaborate with AI professionals to explore the approach of integrating the cognitive computing method with the human-in-the-loop method, either at both system and/or at biological levels (e.g., brain-computer interface).

**The challenges in collaborations**

Over the last fifty years, there have been collaborations between AI and HCI communities. When the development of AI encountered a bottleneck, HCI often provided new research ideas and application scenarios for AI technology such as voice input. AI has brought breakthroughs to HCI technology and elevated HCI to a new development space.

As an emerging approach, HCAI will need collaboration from both communities and the collaboration faces challenges in practice. Research shows that while HCI professionals are challenged by the need to effectively influence the development of AI systems, AI professionals may not fully understand HCAI, and many HCI professionals still join AI projects only *after* requirements are defined, a typical problem when HCI was an emerging field forty years ago. Consequently, recommendations from HCI professionals could be easily ignored by AI professionals [5]. AI professionals often claim that many ease-of-use issues for UI design that HCI could not solve in the past have been solved through AI technology (e.g., voice input). However, studies have shown that the outcomes followed by a technology-driven approach may not be acceptable from the UX perspective [7]. Also, AI and HCI professionals find it challenging to collaborate with each other effectively. Recent studies have shown that HCI professionals do not seem to be prepared to provide effective design support for AI systems due to lack of knowledge of AI, while AI professionals may not fully understand the purpose behind HCI work [5]. There can be no well-integrated process between the two sides without a shared language and process.

**Recommendations for the shared journey of enabling HCAI**

We call for actions below to foster the collaboration between the two communities.

**1) Share a common design philosophy.** In response to the challenges faced by the interdisciplinary communities a few decades ago, William Howell (2011) proposed a "shared philosophy" which integrates human-centered design philosophy with other disciplines [8]. Over the past forty years, the participation from multiple disciplines in the field of HCI for promoting the design philosophy is the embodiment of this model. HCI and AI professionals need to jointly promote HCAI as we did forty years ago for promoting the "human-centered design" approach for PC applications.

**2) Apply an integrated interdisciplinary approach.** To enable HCAI in practice, both communities need to enhance their own methods by leveraging the other. For example, we need to enhance current processes for developing AI systems by incorporating HCI processes and methods, such as iterative prototyping and UX testing, as well as enhancing current software verification/validation methodology by effectively managing evolving machine behavior in AI systems. Recent work on Interactive Machine Learning shows promising [8], whereas the "AI as a material" approach helps HCI professionals improve current HCI design in developing AI systems [10]. Thus, members from diverse disciplines can collaborate to attain shared goals with complementary methods.

**3) Build optimal UX in AI systems.** Some AI professionals assume that AI technologies already make intelligent UI usable (e.g., with voice input) and have little concern about UX. The reality is that we definitely need more UX work when developing AI-based intelligent interaction [4]. HCI professionals can develop new interaction paradigms for usable and natural intelligent UI, and further support the new paradigm of human-AI collaboration. A good example for the collaboration to address the AI black box problem, a collaborative work between AI and HCI leads to more explainable and interpretable AI decision for the drivers of autonomous vehicle through a human-centered XAI approach [11].

**4) Enhance HCI and UX design with AI**. AI technologies have transformed the way for HCI and UX design. For examples, we may use machine learning/algorithm-based approaches to identify insights in user research, design, and UX evaluation. Despite attempts to integrate HCI and AI, HCI designers experience challenges in incorporating ML into UX design and when collaborating with data scientists. HCI and AI communities need to collaborate in developing innovative methods, tools, and processes to help HCI design better innovate with AI.

**5) Design ethical AI collaboratively**. Many AI-related ethical standards are available now. However, the implement of ethical AI is still a challenge in the real world. Some AI professionals view ethical decision-making as another form of technical problem solving, and many also lack formal training in applying ethics to design, so that the community lacks the technical knowledge and solution examples to implement ethical AI. A multidisciplinary approach may help achieve ethical AI. For instance, AI professionals may apply HCI iterative prototyping/testing and behavioral science methods to improve the training and validation of algorithms to minimize algorithmic bias.

**6) Update skillset and knowledge.** While AI professionals should understand HCI, HCI professionals also need to understand AI technology and apply it to facilitate collaboration. Such a mutual understanding from an interdisciplinary perspective will overcome their lack of ability to influence on AI systems as reported today [5]. AI professionals also need to obtain the necessary knowledge from HCI, behavioral and social sciences.

**7) Train the next generation of AI developers and designers.** Over the past forty years, HCI, human factors and psychology have provided an extensive array of professional capabilities, contributing to a mature UX culture. For this to occur for HCAI, new measures at the level of college education are required, including cultivating interdisciplinary skills. This can include providing students with multiple options of a hybrid curriculum of "HCI + AI," or "AI major + social science minor."

**8) Accelerate interdisciplinary research and application.** The development of AI technology itself has benefited from interdisciplinary collaboration. We advocate further collaborative projects for developing AI systems across disciplines and domains. The development of autonomous vehicles is a good example. Many companies are currently investing heavily in developing autonomous vehicles and there are opportunities for collaboration to overcome the challenges that are coming to light as they encounter the real world.

**9) Foster a mature culture of HCAI.** To foster a mature culture of HCAI, we need support through management commitment, organizational culture, optimized development process, design standards and governance, and so on. We firmly believe that a mature HCAI culture will eventually come into being, and history has proven our initial success in promoting the human-centered design philosophy through collaboration in the PC era.

To conclude, although the HCAI approach is in its initial stages, but its influence will ultimately determine our continuing efforts, just as the culture of UX has been jointly developed over the last forty years. We thus find ourselves at a new historical junction, initiating a new journey of collaboration between the AI and HCI communities to enable the HCAI approach.

**Wei Xu** is a Professor at Zhejiang University, China. He received his Ph.D. in Psychology with emphasis on HCI and his M.S. in Computer Science from Miami University in 1997. His research interests include human-AI interaction, HCI, and aviation human factors.

**Marvin Dainoff** is a Professor Emeritus at Miami University. He is past president of the Human Factors and Ergonomics Society. He received his Ph.D. in psychology from University of Rochester in 1969. His research interests include human factors, sociotechnical approach for complex systems, and workplace ergonomics.